\documentclass[floatfix,superscriptaddress,twocolumn,showpacs,prl,amsmath,amssymb]{revtex4-1}
\usepackage{graphicx} 
\usepackage[dvipdfm]{color}
\usepackage{amsmath}
\usepackage{amsfonts}
\usepackage{amssymb}

\usepackage{bm}
\usepackage{txfonts}
\newcommand{\sub}[1]{$_{\mathrm {#1}}$}
\newcommand{\subm}[1]{_{\mathrm {#1}}}

\newcommand{\spsm}[1]{^{\mathrm {#1}}}

\renewcommand{\deg}{^{\circ}}
\newcommand{\Tc}{T\subm{c}}

\newcommand{\Hcc}{H\subm{c2}}
\newcommand{\Hcct}{H\subm{c2}\spsm{\,thm}}

\newcommand{\dash}{^{\prime}}
\newcommand{\tm}[1]{(TMTSF)\sub{2}{#1}}
\newcommand{\tmc}{\tm{ClO\sub{4}}}

\newcommand{\tmx}{\tm{$X$}}

\newcommand{\cstar}{c^{\ast}}
\newcommand{\bstar}{b^{\ast}}
\newcommand{\Rzz}{R_{c^{\ast}}}
\newcommand{\Hp}{H\subm{P}}

\newcommand{\C}{C}
\newcommand{\Ce}{C_e}

\newcommand{\vF}{\bm{v}\subm{F}}
\newcommand{\kF}{\bm{k}\subm{F}}

\newcommand{\vs}{\bm{v}\subm{s}}
\newcommand{\knode}{\bm{k}\subm{F}\spsm{node}}
\newcommand{\vnode}{\bm{v}\subm{F}\spsm{node}}

\begin{document}

\title{Nodal Superconducting Order Parameter and Thermodynamic Phase Diagram of (TMTSF)$_{\bm{2}}$ClO$_{\bm{4}}$}

\author{Shingo~Yonezawa}
\affiliation{Department of Physics, Graduate School of Science, 
Kyoto University, Kyoto 606-8502, Japan}
\author{Yoshiteru~Maeno}
\affiliation{Department of Physics, Graduate School of Science, 
Kyoto University, Kyoto 606-8502, Japan}

\author{Klaus~Bechgaard}
\affiliation{Department~of~Chemistry, Oersted~Institute, Universitetsparken 5, 2100 Copenhagen, Denmark}

\author{Denis~J{\'{e}}rome}
\affiliation{Laboratoire de Physique des Solides (UMR 8502) - Universit{\'{e}} Paris-Sud, 91405 Orsay, France}

\email{yonezawa@scphys.kyoto-u.ac.jp}

\date{\today}

\begin{abstract}
The organic materials \tmx\ are unique unconventional superconductors with archetypal quasi-one-dimensional (Q1D) electronic structures.
Here, based on our comprehensive field-angle-resolved calorimetry of \tmc, we succeeded in mapping the nodal gap structure for the first time in Q1D systems, by discriminating between the Fermi wavevectors and Fermi velocities.
In addition, the thermodynamic phase diagrams of \tmc\ for all principal field directions are obtained.
These findings, providing strong evidence of nodal spin-singlet superconductivity, serves as solid bases for further elucidation of anomalous superconducting phenomena in \tmx.
\end{abstract}

\maketitle

It was in 1980 when the first organic superconductor tetramethyl-tetraselena-fulvalene (TMTSF) salt was discovered.~\cite{Jerome1980,Bechgaard1981}
Since then,  \tmx\ ($X$ = ClO$_{4}$, PF$_6$, etc.) have been widely studied because of their fascinating properties resulting from their archetypal quasi-one-dimensional (Q1D) conductivity and strong electron-electron interactions.~\cite{IshiguroYamajiText,Kuroki2005.JPhysSocJpn.74.1694,LebedText,Note.Band}
Interestingly, the superconducting (SC) phase is located next to a magnetic phase, resembling superconductivity in other unconventional superconductors such as high-$\Tc$ cuprates and pnictides.~\cite{DoironLeyroud2009.PhysRevB.80.214531,Jin2011.Nature.476.73}
Because of this similarity, as well as the simplicity of the electronic structure, investigations of \tmx\ can provide useful guidelines for studies of other unconventional superconductors.

Many studies have revealed unusual SC phenomena in \tmx.
For example, the onset of superconductivity in resistivity is observed even above 4~T when the field is parallel to the $a$ axis, the most conducting direction, or to the $b\dash$ axis, the second-most conducting direction.~\cite{Lee1997,Oh2004,Yonezawa2008.PhysRevLett.100.117002,Yonezawa2008.JPhysSocJpn.77.054712} 
This fact indicates that a certain contribution of superconductivity survives beyond the Pauli limiting field $\mu_0\Hp\sim{}$ 2.3-2.6~T,~\cite{Yonezawa2008.JPhysSocJpn.77.054712} where ordinary singlet pairs would be unstable due to the Zeeman splitting. 
Thus, possibilities of a spin-triplet pairing state or a spatially-modulated spin-singlet pairing state, which is the so-called Fulde-Ferrell-Larkin-Ovchinnikov (FFLO) state, have been discussed.~\cite{Lebed1986.JETPLett.44.114,Lee1997,Oh2004,Yonezawa2008.PhysRevLett.100.117002,Yonezawa2008.JPhysSocJpn.77.054712,Lebed2011.PhysRevLett.101.087004}
In addition, the high-field superconductivity is accompanied by a peculiar anisotropy of the resistivity onset.~\cite{Yonezawa2008.PhysRevLett.100.117002,Yonezawa2008.JPhysSocJpn.77.054712}
However, SC phase diagrams have not been established from thermodynamic measurements. 

For clarification of the origin of such unusual behavior, as well as for identification of the SC mechanism of \tmx, orbital and spin parts of the SC order parameter is essentially important.
Although sign changes on the SC gap is evidenced by the suppression of superconductivity by a tiny amount of non-magnetic impurities,~\cite{Choi1982.PhysRevB.25.6208,Joo2005} details of the gap structure and the SC symmetry are still controversial.~\cite{Takigawa1987,Shinagawa2007,Belin1997}
Theories based on a simple Q1D model with the Fermi surface (FS) consisting of a pair of warped sheets have revealed that a spin-singlet $d$-wave-like state with line nodes is stable when spin fluctuations drive the pairing,~\cite{Duprat2001.EurPhysJB.21.219,Kuroki2005.JPhysSocJpn.74.1694} whereas a spin-triplet $f$-wave-like state with similar line nodes can be stable when charge fluctuations are incorporated.~\cite{Kuroki2005.JPhysSocJpn.74.1694}
The ClO\sub{4} salt have a slightly different FS due to the orientational order of the tetrahedral ClO\sub{4} anions at $T\subm{AO}= 24$~K; this order leads to a folding of the band structure along the $\bstar$ direction and the FS splits into two pairs.~\cite{Pevelen2001EurPhysJB}
A nodeless $d$-wave-like, a nodeless $f$-wave-like, and a nodal $d$-wave-like states have been proposed for such a FS.~\cite{Shimahara2000.PhysRevB.61.R14936,Mizuno2011.PhysicaC.471.49}

\begin{figure}
\begin{center}
\includegraphics[width=7.5cm]{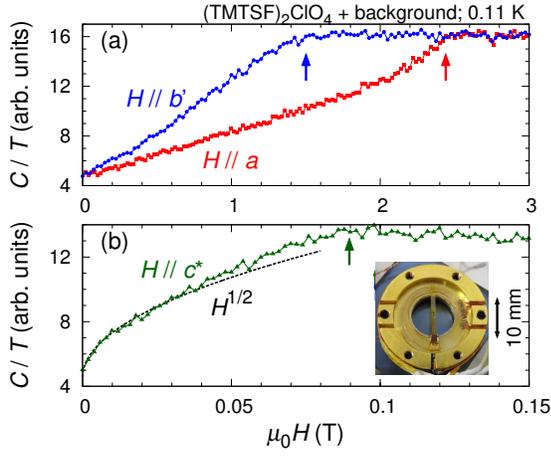}
\end{center}
\caption{
(color online)
Magnetic field dependence of $\C/T$ at 0.11~K for (a) $H\parallel a$ (red squares), $H\parallel b\dash$ (blue circles) and (c) $H\parallel \cstar$.
The arrows indicate the onset $\Hcc$.
The broken curve in (b) illustrates $H^{0.5}$ behavior.
The inset in (b) is a photo of our calorimeter.
\label{fig:H-sweep}}
\end{figure}

To clarify these issues,
we performed field-angle-resolved calorimetry for {\em one} piece of a \tmc\ single crystal.
We used a crystal weighing as low as 76~{$\muup$}g grown by an electrocrystallization technique.~\cite{Yonezawa2011.SM.PRL}
This piece of crystal was previously used in our previous transport study, and was confirmed to be very clean with the mean free path of as large as 1.6~$\muup$m.~\cite{Yonezawa2008.PhysRevLett.100.117002,Yonezawa2008.JPhysSocJpn.77.054712}
We used a ${}^3$He-${}^4$He dilution refrigerator to cool the sample. 
After we cooled the cryostat and sample to 4.2~K, we heated the sample again to 26~K, and cooled it very slowly across $T\subm{AO}$ to 20~K at 4~mK/min, so that the anions order well to be in the ``relaxed'' state.
We newly developed a high-resolution ($\sim 100$~pJ/K at 1~K) calorimeter shown in Fig.~\ref{fig:H-sweep}(b), based on a modification of the ``bath-modulating method''.~\cite{Graebner1989}
The advantage of this technique is that a heater on the sample holder is not necessary; thus the background heat capacity of the sample holder can be minimized. 
Although this background contribution is not subtracted from the data shown here, we have checked that the background is nearly field-independent.~\cite{Yonezawa2011.SM.PRL}
The magnetic field is applied using the vector magnet system.~\cite{Deguchi2004RSI} 
The magnetic field is aligned to the crystalline axes by making use of the anisotropy in the upper critical field $\Hcc$.
The precision and accuracy of the field alignment is approximately 0.1$\deg$.
More details of the experimental procedure will be described elsewhere.~\cite{Yonezawa2011.unpublished}

First we focus on the field dependence of $\C/T$ plotted in Fig.~\ref{fig:H-sweep}.
For $H\parallel \cstar$, i.e. fields perpendicular to the conducting $ab\dash$ plane, $\C(H)/T$ exhibit a $H^{0.5}$ dependence at low temperatures.
This dependence provides strong evidence for a line-node gap.~\cite{Volovik1993.JETPLett.58.469}
In contrast, $\C(H)/T$ for $H\parallel a$, for which the orbital pair breaking is substantially weakened, exhibits a concave-up curvature at low temperatures (Fig.~\ref{fig:H-sweep}(a)) near $\Hcc$. 
Considering the clear change of the spin susceptibility in the SC state,~\cite{Shinagawa2007} we identify this behavior as the Pauli limiting behavior in a spin-singlet superconductor.~\cite{Ichioka2007.PhysRevB.76.064502}

\begin{figure}
\begin{center}
\includegraphics[width=8.5cm]{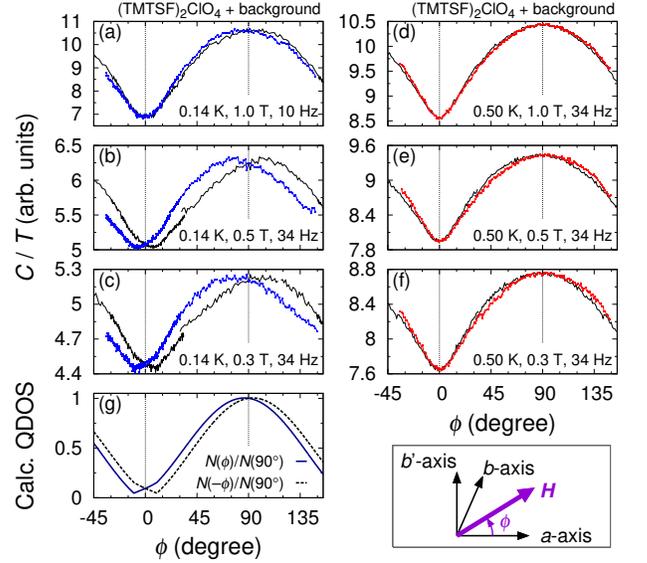}
\end{center}
\caption{
(color online)
Magnetic-field-angle $\phi$ dependence of the heat capacity for fields rotated within the conducting $ab\dash$ plane at (a)--(c) 0.14~K, and (d)--(f) 0.50~K.
For comparison, the same data are also plotted against $-\phi$ with appropriate shifting (the black curves). 
The deviation of the two curves indicates the asymmetry in the $\C(\phi)/T$ curve.
(g) QDOS $N(\phi)/N(90\deg)$ (blue solid curve) and $N(-\phi)/N(90\deg)$ (black broken curve) calculated by eq.~\eqref{eq:phi-dependence} assuming two nodes (n1 and n2) with the parameters $\phi\subm{n1}=-10\deg$ and $\phi\subm{n2}=+10\deg$, $A\subm{n2}/A\subm{n1}=0.3$, 
and $\varGamma=\Hcc(0\deg)/\Hcc(90\deg) = 3.5$.
\label{fig:PS-sweep}
}
\end{figure}

\begin{figure}
\begin{center}
\includegraphics[width=7.5cm]{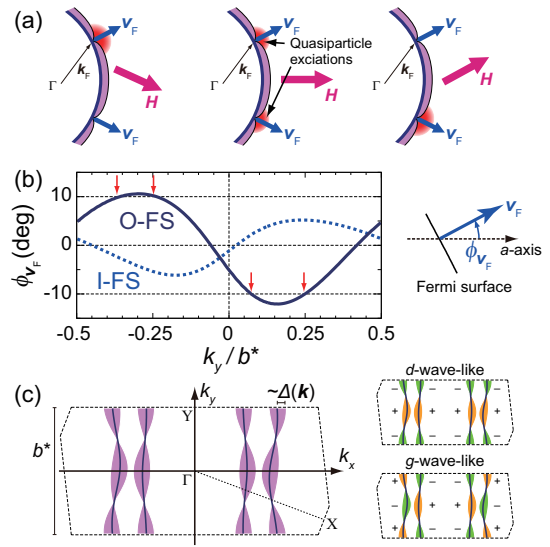}
\end{center}
\caption{
(color online)
(a) Quasiparticle excitation due to the Doppler shift.
When the field is parallel to $\vF$ at a node, the quasiparticle excitation  at this node is reduced.
(b) Dependence of $\phi_{\vF}\equiv \arctan(v_y/v_x)$ on $k_y$ for the outer FS (O-FS; solid curve) and inner FS (I-FS; broken curve). The arrows indicate the points where $\phi_{\vF}$ reaches $\pm10\deg$.
(c) Most plausible gap structure with nodes or zeros at $k_y\sim\pm 0.25b^\ast$. 
Examples of the superconducting states that satisfy the observed nodal structure are shown: the $d$-wave-like state and the $g$-wave-like state. 
\label{fig:FS-vF}
}
\end{figure}

To investigate the nodal structure of the SC gap, we measured the in-plane field-angle dependence of $\C/T$ represented in Fig.~\ref{fig:PS-sweep}.
When the SC gap has a node or a zero at $\knode$,
the quasiparticle (QP) density of states (QDOS) $N$, which is proportional to $\Ce/T$ at low temperatures, varies with the field direction.
Here, $\Ce$ is the electronic heat capacity.
In the low-temperature and low-field limit, such variation of QDOS originates from the ``Doppler shift'' of the QP energy $\delta\omega(\bm{r}, \bm{k})\propto\vs(\bm{r})\cdot\vF(\bm{k})$, where $\vs(\bm{r})$ is the velocity of the supercurrent around a vortex and $\vF(\bm{k})$ is the Fermi velocity.~\cite{Volovik1993.JETPLett.58.469,Vekhter1999}
When $\delta\omega$ is larger than the gap $\Delta(\bm{k})$, QPs with the wavevector $\bm{k}$ are excited.
Because most of the excitation occurs in the vicinity of the nodes, the most important term is $\delta\omega\spsm{node}\propto\vs\cdot\vnode$, where $\vnode\equiv \vF(\knode)$.
When $\bm{H}$ is parallel to $\vnode$, 
$\delta\omega\spsm{node}$ becomes zero (i.e. $\vs\perp\vnode$) because $\vs$ is perpendicular to $\bm{H}$, and QDOS induced by this shift becomes small.~(Fig.~\ref{fig:FS-vF}(a))
Using this idea, we can investigate {\em the directions of $\vF$ at nodal positions}. 
If the condition $\kF\parallel\vF$, is satisfied, the $k$-space nodal direction equals the field direction for which $C/T$ exhibits minimum.~\cite{Sakakibara2007.JPhysSocJpn.76.051004.review}
For Q2D or 3D systems, the assumption $\kF\parallel\vF$ is reasonable as a simple model.
In contrast, for Q1D systems, $\vF$ is {\em not} parallel to $\kF$ even in the simplest model.
Thus, to deduce the nodal position in the $k$-space, information on the band structure is required. 
Another difficulty for Q1D systems is that a large in-plane anisotropy in $\Hcc$ also contributes to the anisotropy in $\Ce$.
Because of these difficulties, $\Ce/T$ oscillation in Q1D systems has been little studied, despite its essential importance in determining the SC gap structure.

\begin{figure}[bth]
\begin{center}
\includegraphics[width=7.5cm]{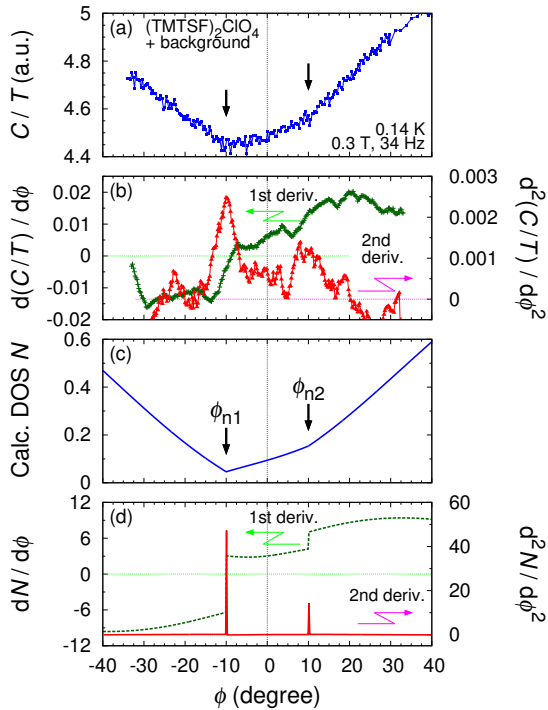}
\end{center}
\caption{
(color online)
(a) Enlarged view of $C(\phi)/T$ at 0.14~K and 0.3~T. The arrows indicate the positions of the small kinks.
(b) Derivatives $\mathrm{d}(C/T)/\mathrm{d}\phi$ (green crosses) and $\mathrm{d}^2(C/T)/\mathrm{d}\phi^2$ (red triangles).
(c) Calculated DOS near $\phi=0$. 
(d) Qualitative behavior of $\mathrm{d}N/\mathrm{d}\phi$ (green broken curve) and $\mathrm{d}^2N/\mathrm{d}\phi^2$ (red solid curve).
\label{fig:PS-sweep2}
}
\end{figure}

Interestingly, we find that the $\C(\phi)/T$ curves of \tmc, where $\phi$ is the azimuthal angle of the field measured from the $a$ axis, become asymmetric with respect to the $a$ axis (e.g. $\C(\phi) > \C(-\phi)$ for $0\deg<\phi<90\deg$) at low temperatures and low fields, as shown in Figs.~\ref{fig:PS-sweep}(a) and (b). 
In contrast, the curves are nearly symmetric at high temperatures or in high fields.
Thus the asymmetry is not due to a misalignment of the field. 
What is more, at 0.14~K, small kinks are observed at around $\phi=\pm10\deg$ as shown in Fig.~\ref{fig:PS-sweep2}(a).
The kink signatures are more obvious in the derivatives (Fig.~\ref{fig:PS-sweep2}(b)):
The step-like behavior in the first derivative and the peaks in the second derivative provide clear indication of the kinks.

The results qualitatively agrees with the theoretical expectation that the specific-heat anomaly due to the gap anisotropy should appear only in the low-temperature and low-field region~\cite{Sakakibara2007.JPhysSocJpn.76.051004.review,Nagai2011.PhysRevB.83.104523}. 
Thus we attribute them to the SC gap anisotropy.
Below, we demonstrate that a simple model based on the Doppler shift reproduces the key features of the unconventional behavior. 
In order to incorporate the large in-plane $\Hcc$ anisotropy into the Doppler-shift mechanism,~\cite{Vekhter1999} we assume that QDOS $N(\phi)$ for $H\ll\Hcc(\phi)$ varies as 
$N(\phi) \propto \sqrt{{H}/{\Hcc(\phi)}}\sum_{n} A_n|\sin(\phi-\phi_n)|$,
where $\phi_n$ is the direction of $\vF$ at the $n$-th node/zero.
To take into account the triclinic band structure, we introduce the nodal-position dependent amplitude $A_n$.
We approximate $\Hcc(\phi)$ to follow the effective mass model:
$\Hcc(\phi) = \Hcc(0\deg)/(\varGamma^2\sin^2\!\phi + \cos^2\!\phi)^{1/2}$,
with $\varGamma\equiv\Hcc(0\deg)/\Hcc(90\deg)$.
Thus $N(\phi)$, which is proportional to $\Ce/T$, should vary as
\begin{align}
N(\phi) \propto &(\varGamma^2\sin^2\!\phi + \cos^2\!\phi)^{1/4} \nonumber\\
&\times\sqrt{\frac{H}{\Hcc(0\deg)}}\sum_{n\mathrm{:nodes/zeros}} A_n|\sin(\phi-\phi_n)|.
\label{eq:phi-dependence}
\end{align}
The calculated $N(\phi)$ with $\phi_n=\pm 10\deg$ plotted in Figs.~\ref{fig:PS-sweep}(g) and \ref{fig:PS-sweep2}(c) well captures the observed unconventional behavior, though we could not perfectly fit the equation to the data.
For a better fitting, additional contributions of vortex cores, thermal excitations, and the Pauli effect should probably be taken into account.
It is worth noting that the behavior of the derivatives is also consistent with the experiment.
We note that a recent theory~\cite{Nagai2011.PhysRevB.83.104523} also attributed the observed calorimetric behavior to the existence of the gap node.

The above analysis indicates that at least there must be a node (or zero) with $\vF$ pointing $\phi=+10\deg$ and another one with $\vF$ pointing $-10\deg$.
To determine the nodal position in the $k$-space, we plot in Fig.~\ref{fig:FS-vF}(b) the $k_y$ dependence of the velocity angle measured from the $a$ axis, $\phi_{\vF}$, based on the tight-binding band structure~\cite{Pevelen2001EurPhysJB,Note.Band}. 
Fig.~\ref{fig:FS-vF}(b) manifests that $|\phi_{\vF}|$ reaches $10\deg$ on the outer FS at $k_y \sim \pm0.25b^\ast, +0.36b^\ast, -0.06b^\ast$, where $b^\ast$ is the size of the 1st Brillouin zone along $k_y$.
Thus, at least, some of the gap nodes or zeros should be located around these positions.
The simplest structure that satisfies this condition is a structure with nodes or zeros running at $k_y = \pm0.25b^\ast$, which is presented in Fig.~\ref{fig:FS-vF}(c).
Leading candidates of the pairing state that satisfies the obtained gap structure are the $d$-wave-like or the $g$-wave like states shown in Fig.~\ref{fig:FS-vF}(c).

\begin{figure}
\begin{center}
\includegraphics[width=8.5cm]{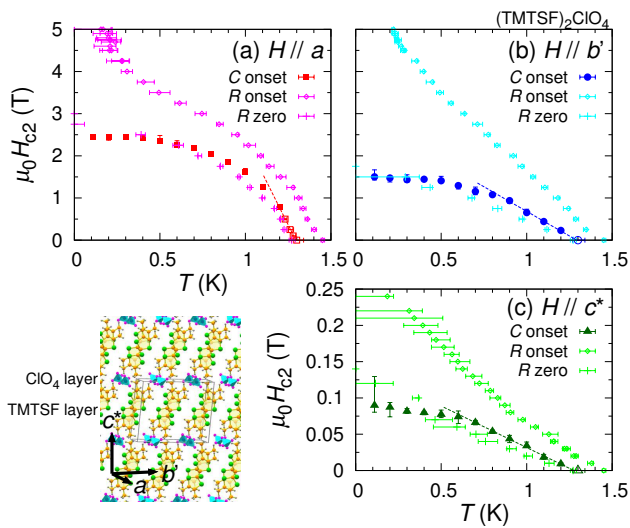}
\end{center}
\caption{
(color online)
Thermodynamic SC phase diagrams of \tmc\ compared with the transport phase diagrams for fields parallel to (a) the $a$ axis, (b) the $b\dash$ axis, and (c) the $\cstar$ axis. 
The closed symbols are obtained from field sweeps and the open symbols from temperature sweeps.
For comparison, we also present the onset temperature of the $\cstar$-axis resistance $\Rzz$ (open diamonds), as well as the temperature where $\Rzz$ becomes zero (crosses), reported in Ref.~\citenum{Yonezawa2008.PhysRevLett.100.117002}.
The broken lines indicate the slope $\mu_0\mathrm{d}\Hcc(T)/\mathrm{d}T$ near $T=\Tc$: $-8.1$~T/K for $H\parallel a$, $-2.3$~T/K for $H\parallel b\dash$, and $-0.11$~T/K for $H\parallel \cstar$. 
The crystalline structure and axes are schematically shown.
\label{fig:phase_diagram}}
\end{figure}

Finally, in Fig.~\ref{fig:phase_diagram}, we present the thermodynamic SC phase diagrams based on the calorimetry compared with transport phase diagrams based on the $\cstar$-axis resistivity.~\cite{Yonezawa2008.PhysRevLett.100.117002,Yonezawa2008.JPhysSocJpn.77.054712}
The thermodynamic upper critical field $\mu_0\Hcct(0) \simeq 2.5$~T for $H\parallel a$ is much smaller than $\mu_0H\subm{c2}\spsm{orb}=-0.73\Tc(\mu_0\mathrm{d}\Hcc/\mathrm{d}T|_{T=\Tc}) \sim 7.7$~T expected for the orbital pair-breaking but agrees with the Pauli-limiting field $\mu_0\Hp\sim 2.3$-$2.6$~T.~\cite{Yonezawa2008.JPhysSocJpn.77.054712} 
This fact again supports a spin-singlet scenario.
Furthermore, the absence of multiple SC phases provides the first proof of a singlet state in the region below $\Hcct$.
Looking again at earlier thermodynamic studies here, one notices that the present $\Hcct(0)$ agrees with the field at which the nuclear-lattice relaxation rate recovers to the normal-state value,~\cite{Shinagawa2007} or with the irreversible field in the torque measurement.~\cite{Oh2004}
Thus, these anomalies reported earlier are now turned out to be due to the thermodynamic SC transition.

Although we used the identical crystal in both the thermodynamic and transport studies, the obtained phase diagrams are quite different.
It is clear that a long-range ordered SC state does not exist above $\Hcct$.
In the region between $\Hcct$ and the resistance onset, however, the observed sharp resistance drop and the anisotropy in the onset temperature reflecting the bulk Fermi surface anisotropy~\cite{Yonezawa2008.PhysRevLett.100.117002,Yonezawa2008.JPhysSocJpn.77.054712} evidence that intrinsic superconductivity robustly contributes to the transport.
The superconductivity above $\Hcct$ should be realized without any noticeable entropy change compared to the normal state.
Thus, it must be characterized either by a fluctuating order parameter or by a static order parameter accompanied by only a tiny change in the density of states.
The true nature of the superconductivity in the high-field region is one of the most intriguing problems to be clarified in future.

In summary, our precise field-angle-resolved calorimetry using a very clean single crystal of \tmc\ provide strong evidence of a spin-singlet nodal superconductivity.
Our result marks the first calorimetric mapping of the gap structure of a Q1D superconductor beyond the constraint previously believed;
this technique is indeed applicable regardless of crystalline and electronic symmetries.
We also revealed the thermodynamic SC phase diagram, which exhibits substantial deviation from the resistivity onset.
The present findings should serve as solid bases for investigation of the pairing mechanism of \tmx, as well as for elucidation of unconventional SC phenomena in this compound.

We acknowledge P. Auban-Senzier, S. Kittaka, and T. Nakamura for their supports; and Y. Matsuda, Y. Nagai, R. Ikeda, C. Bourbonnais, Y. Fuseya, A. Kobayashi, Y. Suzumura, A. Schofield, S. Brown, H. Shimahara, H. Aizawa for useful discussions.
We also acknowledge KOA Corporation, TORAY Advanced Film, and TORAY for providing us with their products for the calorimeter.
This work is supported by a Grant-in-Aid for the Global COE ``The Next Generation of Physics, Spun from Universality and Emergence'' and by Grants-in-Aids for Scientific Research (KAKENHI 21110516, 21740253, 23540407, and 23110715) from MEXT and JSPS.

\bibliography{./MainPaper,%
D:/cygwin/home/Owner/SSP/paper/string,%
D:/cygwin/home/Owner/SSP/paper/TMTSF,%
D:/cygwin/home/Owner/SSP/paper/textbook,%
D:/cygwin/home/Owner/SSP/paper/FFLO,%
D:/cygwin/home/Owner/SSP/paper/CeCoIn5,%
D:/cygwin/home/Owner/SSP/paper/superconductors,%
D:/cygwin/home/Owner/SSP/paper/SC,%
D:/cygwin/home/Owner/SSP/paper/Sr2RuO4,%
D:/cygwin/home/Owner/SSP/paper/high-Tc,%
D:/cygwin/home/Owner/SSP/paper/measurement_technique%
}

\begin{thebibliography}{33}
\bibitem [{Not()}]{Note.Band}%
  \BibitemOpen
  \bibinfo {note} {We note here that a single-electron band picture has been
  proved to be quite reasonable for
  (TMTSF)$_X$~\cite{IshiguroYamajiText,LebedText} in spite of strong
  electron-electron correlation. This is in clear contrast to e.g.\
  {$f$}-electron heavy-Fermion systems.}\BibitemShut {Stop}%
\end{thebibliography}

\clearpage

\begin{widetext}

\begin{center}
{\large \bfseries Supplemental Material} 
\end{center}

\section{Sample}

In the present study, we used a single crystal weighing as low as 76~{$\muup$}g shown in Fig.~\ref{fig:T-step}(a),
whose heat capacity at temperatures slightly above $\Tc$ is estimated to be approximately 3~nJ/K based on the result in the previous study.~\cite{Garoche1982.JPhysLett.43.L147}
It should be noted that this piece of crystal was previously used in our previous transport study~\cite{Yonezawa2008.PhysRevLett.100.117002};
we removed the electrodes from the crystal and used it for the present calorimetry.
This allows us to compare superconducting properties of the identical sample.
A small amount of gold used for the electrode is left on the surface of the crystal as seen in the photo.
The contribution of this gold to the heat capacity is estimated to be on the order of 1~pJ/K at 1~K and is negligible.

We note here that \tmx\ salts are particularly stable in the air and at room temperature over many years and don't suffer from aging. 
Indeed, recently published transport work has been performed on excellent 30 years old crystals.~\cite{DoironLeyroud2009.PhysRevB.80.214531} 
Thus, the oldness of our sample is not a problem, although a few years have passed since we used the crystal for the previous transport study. We also note that thermal cycling of \tmx\ crystals are not reported to affect bulk measurements, although transport measurements are known to be affected by temperature cycling mainly by micro-cracks within crystals.

\section{Temperature Dependence of the Heat Capacity and Background}

Our high-resolution ($\sim 100$~pJ/K at 1~K) calorimeter used in this study is shown in Fig.~\ref{fig:T-step}(a). 
This calorimeter is developed based on a modification of the ``bath-modulating method'':~\cite{Graebner1989}
For our apparatus, small thick-film resistors are used for thermometers, though thermocouples are used in Ref.~\cite{Graebner1989}.
The advantage of the bath-modulating technique is that one does not need to place a heater on the sample stage; thus the background heat capacity of the sample holder can be minimized. 
With this technique, sensitivity high enough to measure the heat capacity of one single crystal of \tmc\ is achieved, although we can only obtain the relative value of the heat capacity. 
Depending on the temperature and field conditions, we chose the temperature modulation frequency as 34~Hz or 10~Hz,
which is twice of the heater-current frequency.
We adopted the temperature modulation amplitude of 1.5\%-rms and 3\%-rms of the base temperature for the temperature and field sweeps,
while we used 6\%-rms for the field-angle sweeps in order to improve the signal-to-noise ratio.
More details of the measurement will be described elsewhere.

The heat-capacity data presented in this report contains the contributions from both the sample and the background. 
The background contribution, which is represented in Fig.~\ref{fig:T-step}(b), is dominated by the heat capacity of the thermometer and that of the Apiezon N Grease used to fix the sample to the calorimeter. 
Because our single crystal weighs only 76~$\muup$g,
the background contribution is affected even by a tiny amount of the grease.
The amount of the grease used in the measurement cannot be exactly known.
Therefore, the subtraction of the background contribution is not thoroughly accurate.

\begin{figure*}[bth]
\begin{center}
\includegraphics[width=15.0cm]{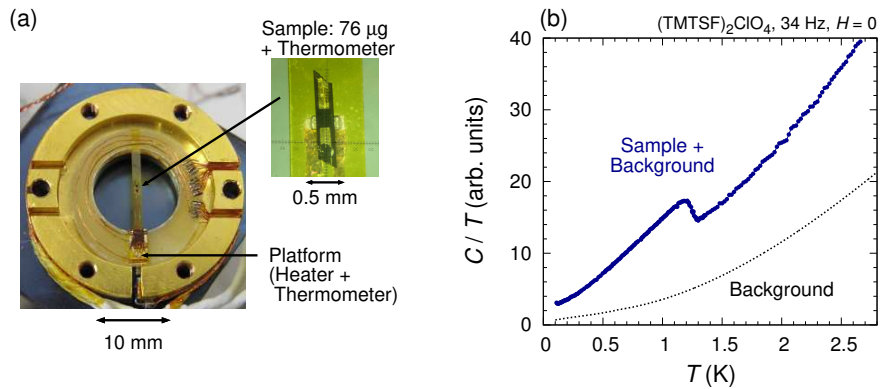}
\end{center}
\caption{Descriptions of our experimental setup.
(a) Photos of the calorimeter and the sample. We used a very clean single crystal of \tmc\ weighing 76 $\muup$g. We note that this crystal is identical to the one used in our previous transport studies.~\cite{Yonezawa2008.PhysRevLett.100.117002,Yonezawa2008.JPhysSocJpn.77.054712}
(b) Temperature dependence of the total (i.e. sample + background) heat capacity (solid circles), compared with the background heat capacity (broken curve). Note that the background contribution is not thoroughly accurate as described in the text.
\label{fig:T-step}
}
\end{figure*}

\section{Field Dependence of the Background}

In order to prove that the background contribution have little influence in the field-amplitude sweeps and field-angle sweeps, we compare the data with the separately measured background data in Fig.~\ref{fig:compare_background}.
Although the actual background may be different by a few tens of percent, its field dependence should be quite similar to the data presented here.
As is clear in this figure, the field-amplitude and field-direction dependences of the background contribution are negligibly small.
Therefore, it is certain that our conclusion of the Letter is not at all affected by the background contribution.

\begin{figure*}[h]
\begin{center}
\includegraphics[width=15.0cm]{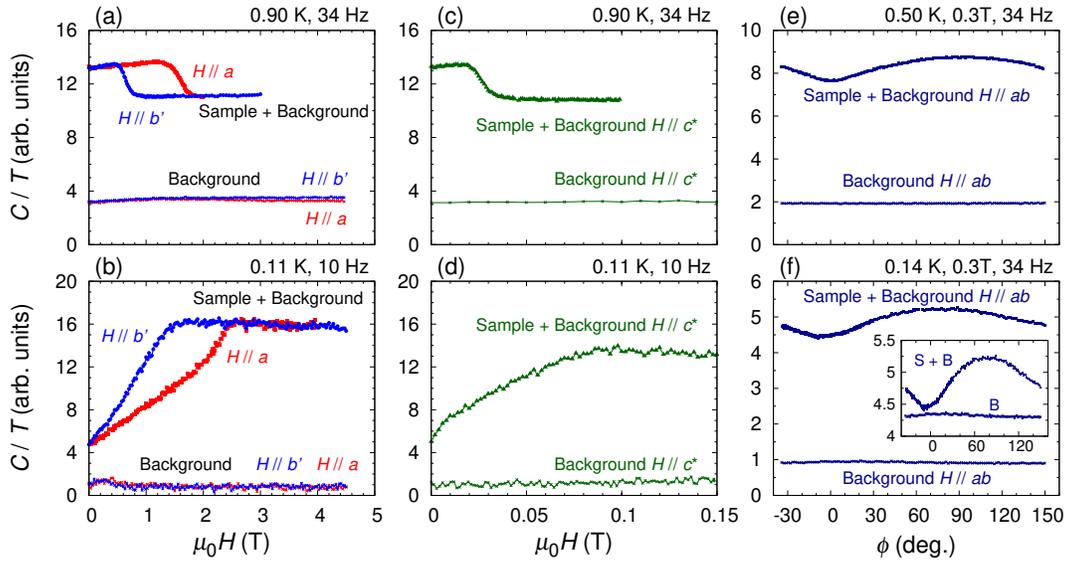}
\end{center}
\caption{Comparison of the experimental data with the separately measured background.
In all cases, the background does not affect the field dependence of the total heat capacity at all.
(a)--(b) Field-sweep data for $H \parallel a$ and $H\parallel b\dash$ at 0.90~K and 0.11~K.
(c)--(d) Field-sweep data for $H \parallel \cstar$ at 0.90~K and 0.11~K.
(e)--(f) In-plane field-direction $\phi$ dependence at 0.50~K and 0.14~K.
\label{fig:compare_background}
}
\end{figure*}

\end{widetext}

\end{document}